\documentclass[aps,floatfix,superscriptaddress,showpacs,showkeys]{revtex4}
\usepackage{amssymb,amsmath,amsfonts,latexsym,graphicx,epsfig}

\usepackage{color}
\usepackage{titlesec}
\titleformat{\section}{\large\bfseries}{\thesection}{1em}{}

\newcommand{\bea}{\begin{eqnarray}}
\newcommand{\ena}{\end{eqnarray}}
\newcommand{\nn}{\nonumber\\}

\newcommand{\be}{\begin{equation}}
\newcommand{\en}{\end{equation}}
\newcommand{\ed}{\end{document}}

\newcommand{\la}{\langle}
\newcommand{\ra}{\rangle}

\newcommand{\Tr}{\mbox{\rm{tr}}}
\newcommand{\oone}{\hbox{$1\kern-2.5pt\hbox{\rm l}$}}
\newcommand{\MM}{\hbox{$\kern2.5pt\vrule height6.7pt\kern-0.5pt{\rm M}$}}
\newcommand{\Jpsi}{\ensuremath{J\!/\!\psi}}

\usepackage{tabularx}
\newcolumntype{R}{>{\raggedleft\let\newline\\\arraybackslash\hspace{0pt}$}X<{$}}
\usepackage{float}
\newcommand{\tH}[1]{|\widehat{H}_{#1}|^2}

\begin{document}

\hfill MITP/17-036 (Mainz)

\title{
Theoretical description of the decays
\boldmath{$\Lambda_b \to  \Lambda^{(\ast)}(\frac12^\pm,\frac32^\pm) + J/\psi$} }

\author{Thomas Gutsche}
\affiliation{Institut f\"ur Theoretische Physik, Universit\"at T\"ubingen,\\
Kepler Center for Astro and Particle Physics,\\
Auf der Morgenstelle 14, D-72076, T\"ubingen, Germany}

\author{Mikhail A. Ivanov}
\affiliation{Bogoliubov Laboratory of Theoretical Physics, \\
Joint Institute for Nuclear Research, 141980 Dubna, Russia}

\author{J\"{u}rgen G. K\"{o}rner}
\affiliation{PRISMA Cluster of Excellence, Institut f\"{u}r Physik,
Johannes Gutenberg-Universit\"{a}t, \\
D-55099 Mainz, Germany}

\author{Valery~E.~Lyubovitskij}
\affiliation{Institut f\"ur Theoretische Physik, Universit\"at T\"ubingen,\\
Kepler Center for Astro and Particle Physics,\\
Auf der Morgenstelle 14, D-72076, T\"ubingen, Germany}
\affiliation{Departamento de F\'\i sica y Centro Cient\'\i fico
Tecnol\'ogico de Valpara\'\i so (CCTVal), Universidad T\'ecnica
Federico Santa Mar\'\i a, Casilla 110-V, Valpara\'\i so, Chile}
\affiliation{Department of Physics, Tomsk State University,
634050 Tomsk, Russia}
\affiliation{Laboratory of Particle Physics, Mathematical Physics Department,
Tomsk Polytechnic University, 634050 Tomsk, Russia}

\author{Vladimir V. Lyubushkin}
\affiliation{Dzhelepov Laboratory of Nuclear Problems, \\
Joint Institute for Nuclear Research, 141980 Dubna, Russia}

\author{Pietro Santorelli}
\affiliation{Dipartimento di Fisica, Universit\`a di Napoli
Federico II, Complesso Universitario di Monte S. Angelo,
Via Cintia, Edificio 6, 80126 Napoli, Italy}
\affiliation{Istituto Nazionale di Fisica Nucleare, Sezione di
Napoli, 80126 Napoli, Italy}


\begin{abstract}

We calculate the invariant and helicity amplitudes for the transitions
$\Lambda_b~\to~\Lambda^{(\ast)}(J^P)~+~J/\psi$, where
the $\Lambda^{(\ast)}(J^P)$ are  $\Lambda(sud)$-type ground and excited states
with $J^P$ quantum numbers
$J^P=\frac12^{\pm},\frac32^{\pm}$. The calculations are performed in
the framework of a covariant confined quark model previously developed by us.
We find that the values of the helicity amplitudes for
the $\Lambda^\ast(1520,\,\frac32^-)$ and the $\Lambda^\ast(1890,\,\frac32^+)$
are suppressed compared with those for the
ground state $\Lambda(1116,\,\frac12^+)$ and the excited state
$\Lambda^\ast(1405,\,\frac12^-)$.
This analysis is important  for the identification of
the hidden charm pentaquark states $P_c^+(4380)$ and $P_c^+(4450)$ which
were discovered in the decay chain
$\Lambda_b^0~\to~P_c^+(~\to~p~J/\psi)~+~K^- $ because the cascade decay chain
$\Lambda_b~\to~\Lambda^\ast(\frac32^\pm)(~\to~pK^-)~+~J/\psi$
involves the same final state.

\end{abstract}


\maketitle

\newpage

\section{Introduction}
\label{sec:intro}

Recently the LHCb Collaboration has performed an angular
analysis of the decay $\Lambda_b \to \Lambda^{(*)} + J/\psi$,
where the $\Lambda_{b}$'s are produced in $pp$ collisions
at $\sqrt{s} = 7$ TeV at the LHC (CERN)~\cite{Aaij:2013oxa}.
They reported on the measurement of the relative magnitude of the helicity
amplitudes in the decay $\Lambda_b \to \Lambda^{(*)} + J/\psi$ by a fit to
several asymmetry parameters in the cascade decay distribution
$\Lambda_b \to \Lambda(\to p\pi^-) + J/\psi (\to \ell^{+}\ell^{-})$ and
$\Lambda_b \to \Lambda^\ast(\to pK^-) + J/\psi (\to \ell^{+}\ell^{-})$. 

In an earlier paper~\cite{Gutsche:2013oea}  we have performed a detailed
analysis of the decay process $\Lambda_b \to \Lambda + J/\psi$ within
a  covariant quark model.  We have worked out two variants of the threefold
joint angular decay distributions in the cascade decay
$\Lambda_b\to \Lambda(\to p\pi^-)\,+\,J/\psi(\to\ell^+\ell^-)$ for polarized
and unpolarized $\Lambda_{b}$ decays. We have further listed results on
helicity amplitudes which determine
the rate and the asymmetry
parameters in the decay processes
$\Lambda_b \to \Lambda(\to p\pi^-)\,+\,J/\psi$ and
$\Lambda_b \to \Lambda(\to p\pi^-)\,+\,\psi(2S)$.

In this paper we calculate the corresponding invariant and helicity
amplitudes in
the transitions $\Lambda_b~\to~\Lambda^{(\ast)}(J^P)~+~J/\psi$ where the
$\Lambda^{(\ast)}(J^P)$ are $\Lambda$-type $(sud)$ ground and excited states with
$J^P$ quantum numbers $J^P=\frac12^{\pm},\frac32^{\pm}$. The calculations are
performed in
the framework of our covariant confined quark model developed previously by us.
We find that the values of the helicity amplitudes for the
$\Lambda_b \to \Lambda^\ast(1520,\,\frac32^-),\Lambda^\ast(1890,\,\frac32^+)$
transitions are suppressed compared with those for the
transitions to the ground state
$\Lambda(1116,\,\frac12^+)$ also calculated in~\cite{Gutsche:2013oea}
and the excited state $\Lambda^\ast(1405,\,\frac12^-)$.
This analysis is important for the identification of
the hidden charm pentaquark states $P_c^+(4380)$ and $P_c^+(4450)$ since the
cascade decay
$\Lambda_b~\to~\Lambda^\ast(\frac12^-,\frac32^\pm)(~\to~pK^-)~+~J/\psi$
involves the same final  states as the decay
$\Lambda_b^0~\to~P_c^+(~\to~p~J/\psi)~+~K^- $. The subject of the hidden
charm pentaquark states has been
intensively discussed in the literature (see e.g.
\cite{Roca:2015tea}-\cite{Bayar:2016ftu}).

Our paper is structured as follows.
In Sec.~\ref{sec:definition}, we give explicit expressions for the hadronic
matrix elements $\la \Lambda_2 | \bar s O^\mu b | \Lambda_1 \ra$
in terms of dimensionless invariant form factors $F^{V/A}_i(q^2)$.
The corresponding vector and axial helicity amplitudes are linearly related
to the invariant form factors. The linear relations are explicitly
calculated and listed.
The helicity amplitudes are the basic building blocks in the calculation of
the rate and in the
construction of the full angular decay distributions for
the cascade decays.

In Sec.~\ref{sec:model_calc}, we construct local interpolating three-quark
currents
corresponding to the $\Lambda^{(\ast)}$ states with parity
$J^P=\frac12^\pm,\frac32^\pm$. We then use nonlocal variants of the local
interpolating currents to
evaluate all invariant amplitudes in the framework of the covariant
confined quark model.
In Sec.~\ref{sec:numerics}, we give numerical results
for the normalized helicity amplitudes and branching ratios.
Finally, in Sec.~\ref{sec:summary}, we summarize our findings.


\section{The decays
  \boldmath{$\Lambda_b \to \Lambda^{(\ast)}(\frac12^\pm,\frac32^\pm) + J/\psi$}
: matrix element  and helicity amplitudes}
\label{sec:definition}


The matrix element of the exclusive decay
$\Lambda_1(p_1,\lambda_1)\to \Lambda_2(p_2,\lambda_2)\,+\,V(q,\lambda_V)$
is defined by (in the present application the vector meson label $V$
stands for the $J/\Psi$)

\be
M(\Lambda_1\to \Lambda_2 + V) =
\frac{G_F}{\sqrt{2}} \, V_{cb} \, V^\ast_{cs} \, C_{\rm eff} \,
f_V \, M_V \, \la \Lambda_2 | \bar s O_\mu b | \Lambda_1 \ra \,
\epsilon^{\dagger\,\mu}(\lambda_V) \,,
\label{eq:matr_LbLJ}
\en
where $M_V$ and $f_V$ are the mass and the leptonic decay constant of
the vector meson $V$,  $O^\mu = \gamma^\mu (1 - \gamma^5)$ and
$|V_{cb}| = 0.0406$ and $|V^\ast_{cs}| = 0.974642$ are
Cabibbo-Kabayashi-Maskawa (CKM) matrix elements.
The coefficient $C_{\rm eff}$ stands for the combination of Wilson coefficients
\be
C_{\rm eff}= C_1 + C_3 + C_5
+ \xi \Big(C_2 + C_4 + C_6\Big) \,.
\label{eq:Wilson_eff}
\en
The  color factor $\xi=1/N_c$ will be set to zero such that we only keep 
the leading term in the  $1/N_c-$expansion.
We take the numerical values of the Wilson coefficients
from~\cite{Altmannshofer:2008dz}:
\be
   C_1 = - 0.257\,, \quad
   C_2 = 1.009\,,   \quad
   C_3 = - 0.005\,, \quad
   C_4 = - 0.078\,, \quad
   C_5 \simeq  0\,, \quad
   C_6 = 0.001 \,.
\label{eq:Wilson}
\en

The hadronic matrix element $\la \Lambda_2 | \bar s O^\mu b | \Lambda_1 \ra$
is expressed in terms of six and eight, respectively, dimensionless
invariant form factors $F^{V/A}_i(q^2)$ viz.

\noindent
Transition $\frac{1}{2}^+ \to \frac{1}{2}^+$\,:
\bea
\la \Lambda_2 | \bar s \gamma_\mu b | \Lambda_1 \ra &=&
\bar u(p_2,s_2)
\Big[ \gamma_\mu F_1^V(q^2)
    - i \sigma_{\mu\nu} \frac{q_\nu}{M_1} F_2^V(q^2)
    + \frac{q_\mu}{M_1} F_3^V(q^2)
    \Big]
    u(p_1,s_1)
\nn
\la \Lambda_2 |\bar s\gamma_\mu\gamma_5b  | \Lambda_1 \ra &=&
\bar u(p_2,s_2)
\Big[ \gamma_\mu F_1^A(q^2)
    - i \sigma_{\mu\nu} \frac{q_\nu}{M_1} F_2^A(q^2)
    + \frac{q_\mu}{M_1} F_3^A(q^2)
\Big] \gamma_5 u(p_1,s_1)
\nonumber
\ena

\noindent
Transition $\frac{1}{2}^+ \to \frac{1}{2}^-$\,:
\bea
\la \Lambda_2 |\bar s \gamma_\mu b | \Lambda_1 \ra &=&
\bar u(p_2,s_2)
\Big[ \gamma_\mu F_1^V(q^2)
    - i \sigma_{\mu\nu} \frac{q_\nu}{M_1} F_2^V(q^2)
    + \frac{q_\mu}{M_1} F_3^V(q^2)
\Big]\gamma_5 u(p_1,s_1)
\nn
\la \Lambda_2 |\bar s \gamma_\mu\gamma_5b | \Lambda_1 \ra &=&
\bar u(p_2,s_2)
\Big[ \gamma_\mu F_1^A(q^2)
    - i \sigma_{\mu\nu} \frac{q_\nu}{M_1} F_2^A(q^2)
    + \frac{q_\mu}{M_1} F_3^A(q^2)
\Big]
u(p_1,s_1)
\nonumber
\ena

\noindent
Transition $\frac{1}{2}^+ \to \frac{3}{2}^+$\,:
\bea
\la \Lambda_2^\ast |\bar s \gamma_\mu b| \Lambda_1 \ra &=&
\bar u^\alpha(p_2,s_2)
\Big[ g_{\alpha\mu} F_1^V(q^2)
    + \gamma_\mu \frac{p_{1\alpha}}{M_1} F_2^V(q^2)
+ \frac{p_{1\alpha} p_{2\mu}}{M_1^2} F_3^V(q^2)
    + \frac{p_{1\alpha} q_\mu}{M_1^2}    F_4^V(q^2)
\Big] \gamma_5
u(p_1,s_1)
\nn
\la \Lambda_2^\ast |\bar s \gamma_\mu\gamma_5b | \Lambda_1 \ra &=&
\bar u^\alpha(p_2,s_2)
\Big[ g_{\alpha\mu} F_1^A(q^2)
    + \gamma_\mu \frac{p_{1\alpha}}{M_1} F_2^A(q^2)
+ \frac{p_{1\alpha} p_{2\mu}}{M_1^2} F_3^A(q^2)
    + \frac{p_{1\alpha} q_\mu}{M_1^2}    F_4^A(q^2)
\Big] u(p_1,s_1)
\nonumber
\ena

\noindent
Transition $\frac{1}{2}^+ \to \frac{3}{2}^-$\,:

\bea
\la \Lambda_2^\ast |\bar s \gamma_\mu b | \Lambda_1 \ra &=& \bar
u^\alpha(p_2,s_2)
\Big[ g_{\alpha\mu} F_1^V(q^2)
    + \gamma_\mu \frac{p_{1\alpha}}{M_1} F_2^V(q^2)
+ \frac{p_{1\alpha} p_{2\mu}}{M_1^2} F_3^V(q^2)
    + \frac{p_{1\alpha} q_\mu}{M_1^2}    F_4^V(q^2)
\Big]  u(p_1,s_1)
\nn
\la \Lambda_2^\ast |\bar s  \gamma_\mu\gamma_5 b | \Lambda_1 \ra &=&
\bar u^\alpha(p_2,s_2)
\Big[ g_{\alpha\mu} F_1^A(q^2)
    + \gamma_\mu \frac{p_{1\alpha}}{M_1} F_2^A(q^2)
+ \frac{p_{1\alpha} p_{2\mu}}{M_1^2} F_3^A(q^2)
    + \frac{p_{1\alpha} q_\mu}{M_1^2}    F_4^A(q^2)
\Big] \gamma_5 u(p_1,s_1)
\nonumber
\ena
where
$\sigma_{\mu\nu} = (i/2) (\gamma_\mu \gamma_\nu - \gamma_\nu \gamma_\mu)$
and all $\gamma$ matrices are defined as in the text book by Bjorken-Drell.
We use the same notation for the form factors $F^{V/A}_i$ in all
transitions even though their numerical values differ.
For completeness we have kept the form factors $F_3^{V/A}$
in the $\frac12^+\to\frac12^\pm$ transitions and $F_4^{V/A}$
in the $\frac12^+\to\frac32^\pm$ transitions
although they do not contribute
to the decay $\Lambda_b \to \Lambda^{(\ast)} + J/\psi$ since
$q_\mu \, \epsilon_V^\mu  = 0$.

Next we express the vector and axial helicity amplitudes
$H_{\lambda_2\lambda_V}$ in terms of the invariant form factors
$F_i^{V/A}$,  where $\lambda_V = \pm 1, 0$ and
$\lambda_2 = \pm 1/2, \pm 3/2$ are  the helicity components of the vector
meson $V$
and the daughter baryon $\Lambda_2$, respectively. Note again that the time-component helicity
amplitudes $H^{V,A}_{\lambda_2\,t}$ do not
contribute since the $J/\psi$ is a spin 1 meson.
We need to calculate the expressions
\be
H_{\lambda_2\lambda_V} =
\la \Lambda_2(p_2,\lambda_2) |\bar s O_\mu b | \Lambda_1(p_1,\lambda_1) \ra
\epsilon^{\dagger\,\mu}(\lambda_V)
=  H_{\lambda_2\lambda_V}^V - H_{\lambda_2\lambda_V}^A \,,
\en
where we split the  helicity amplitudes into their vector and axial parts.
We shall work in the rest frame of the parent baryon $\Lambda_1$ with the
daughter baryon $\Lambda_2$
moving in the positive $z$-direction:
$p_1 = (M_1, \vec{\bf 0})$, $p_2 = (E_2, 0, 0, |{\bf p}_2|)$ and
$q = (q_0, 0, 0, - |{\bf p}_2|)$. In this case
$\lambda_1 = \lambda_2 - \lambda_V$.
Following Ref.~\cite{Faessler:2009xn} one has\\

\noindent
Transition $\frac12^+ \to \frac12^+$\,:
$H^V_{-\lambda_2,-\lambda_V} = + H^V_{\lambda_2,\lambda_V}$ and
$H^A_{-\lambda_2,-\lambda_V} = - H^A_{\lambda_2,\lambda_V}$\,.

\[
\begin{array}{lcrlcl}
H_{\frac12 t}^V &=& \sqrt{Q_+/q^2} \,
\Big( F_1^V M_- + F_3^V \frac{q^2}{M_1} \Big) \qquad
&\qquad
H_{\frac12 t}^A &=& \sqrt{Q_-/q^2} \,
\Big( F_1^A M_+ - F_3^A \frac{q^2}{M_1} \Big)
\\[1.1ex]
H_{\frac12 0}^V &=& \sqrt{Q_-/q^2} \,
\Big( F_1^V M_+ + F_2^V \frac{q^2}{M_1} \Big) \qquad
& \qquad
H_{\frac12 0}^A &=& \sqrt{Q_+/q^2} \,
\Big( F_1^A M_- - F_2^A \frac{q^2}{M_1} \Big)
\\[1.1ex]
H_{\frac12 1}^V &=& \sqrt{2Q_-}
\Big( - F_1^V - F_2^V \frac{M_+}{M_1} \Big) \qquad
& \qquad
H_{\frac12 1}^A &=& \sqrt{2Q_+} \,
\Big( - F_1^A + F_2^A \frac{M_-}{M_1} \Big)
\\
\end{array}
\]

\noindent
Transition $\frac12^+ \to \frac12^-$\,:
$H^V_{-\lambda_2,-\lambda_V} = - H^V_{\lambda_2,\lambda_V}$ and
$H^A_{-\lambda_2,-\lambda_V} = + H^A_{\lambda_2,\lambda_V}$\,.

\[
\begin{array}{lcrlcl}
H_{\frac12 t}^V &=& \sqrt{Q_-/q^2} \,
\Big( F_1^V M_+ - F_3^V \frac{q^2}{M_1} \Big) \qquad
&\qquad
H_{\frac12 t}^A &=& \sqrt{Q_+/q^2} \,
\Big( F_1^A M_- + F_3^A \frac{q^2}{M_1} \Big)
\\[1.1ex]
H_{\frac12 0}^V &=& \sqrt{Q_+/q^2} \,
\Big( F_1^V M_- - F_2^V \frac{q^2}{M_1} \Big) \qquad
& \qquad
H_{\frac12 0}^A &=& \sqrt{Q_-/q^2} \,
\Big( F_1^A M_+ + F_2^A \frac{q^2}{M_1} \Big)
\\[1.1ex]
H_{\frac12 1}^V &=& \sqrt{2Q_+}
\Big( - F_1^V + F_2^V \frac{M_-}{M_1} \Big) \qquad
& \qquad
H_{\frac{1}{2}1}^A &=& \sqrt{2Q_-} \,
\Big( - F_1^A - F_2^A \frac{M_+}{M_1} \Big)
\\
\end{array}
\]

Transition $\frac12^+ \to \frac32^+$\,:
$H^V_{-\lambda_2,-\lambda_V} = - H^V_{\lambda_2,\lambda_V}$ and
$H^A_{-\lambda_2,-\lambda_V} = +  H^A_{\lambda_2,\lambda_V}$.

\bea
H_{\frac12 t}^V &=& - \sqrt{\frac23\cdot \frac{Q_+}{q^2}} \,
\frac{Q_-}{2M_1M_2}
\Big( F_1^V M_1 - F_2^V M_+ + F_3^V \frac{M_+M_--q^2}{2M_1}
+ F_4^V \frac{q^2}{M_1} \Big)
\nn[1.1ex]
H_{\frac12 0}^V &=& - \sqrt{\frac23\cdot \frac{Q_-}{q^2}} \,
\Big( F_1^V \frac{M_+M_--q^2}{2M_2} - F_2^V \frac{Q_+ M_-}{2M_1M_2}
+ F_3^V \frac{|{\bf p_2}|^2}{M_2} \Big)
\nn[1.1ex]
H_{\frac12 1}^V &=& \sqrt{\frac{Q_-}{3}} \,
\Big( F_1^V - F_2^V \frac{Q_+}{M_1M_2} \Big)
\qquad
H_{\frac32 1}^V = -  \, \sqrt{Q_-} \, F_1^V
\nn[2ex]
H_{\frac12 t}^A &=& \sqrt{\frac23\cdot \frac{Q_-}{q^2} }
\frac{Q_+}{2M_1M_2}
 \Big( F_1^A M_1 + F_2^A M_- + F_3^A \frac{M_+M_--q^2}{2M_1}
+ F_4^A \frac{q^2}{M_1}\Big)
\nn[1.1ex]
H_{\frac12 0}^A &=&  \sqrt{\frac23\cdot \frac{Q_+}{q^2} }
\Big( F_1^A \frac{M_+M_--q^2}{2M_2} + F_2^A \frac{Q_-M_+}{2M_1M_2}
+ F_3^A  \frac{|{\bf p_2}|^2}{M_2}  \Big)
\nn[1.1ex]
H_{\frac12 1}^A &=& \sqrt{\frac{Q_+}{3}}
\Big( F_1^A - F_2^A \frac{Q_-}{M_1M_2} \Big)
\qquad
H_{\frac{3}{2}1}^A = \, \sqrt{Q_+} F_1^A
\nonumber
\ena

Transition $\frac12^+ \to \frac32^-$\,:
$H^V_{-\lambda_2,-\lambda_V} = + H^V_{\lambda_2,\lambda_V}$
and
$H^A_{-\lambda_2,-\lambda_V} = - H^A_{\lambda_2,\lambda_V}$.

\bea
H_{\frac12 t}^V &=& \sqrt{\frac23\cdot \frac{Q_-}{q^2}}
\frac{Q_+}{2M_1M_2}
 \Big( F_1^V M_1 + F_2^V M_- + F_3^V \frac{M_+M_--q^2}{2M_1}
+ F_4^V \frac{q^2}{M_1}\Big)
\nn[1.1ex]
H_{\frac12 0}^V &=&  \sqrt{\frac23\cdot \frac{Q_+}{q^2} }
\Big( F_1^V \frac{M_+M_--q^2}{2M_2} + F_2^V \frac{Q_- M_+}{2M_1M_2}
+ F_3^V  \frac{{|\bf p_2}|^2}{M_2}  \Big)
\nn[1.1ex]
H_{\frac12 1}^V &=& \sqrt{ \frac{Q_+}{3} }
\Big( F_1^V - F_2^V \frac{Q_-}{M_1M_2} \Big)
\qquad
H_{\frac32 1}^V = \, \sqrt{Q_+} F_1^V
\nn[2mm]
H_{\frac{1}{2}t}^A &=& - \sqrt{\frac23\cdot \frac{Q_+}{q^2}} \,
\frac{Q_-}{2M_1M_2} \Big( F_1^A M_1 - F_2^A M_+ + F_3^A \frac{M_+M_--q^2}{2M_1}
+ F_4^A \frac{q^2}{M_1} \Big)
\nn[1.1ex]
H_{\frac{1}{2}0}^A &=& - \sqrt{\frac23\cdot \frac{Q_-}{q^2}} \,
\Big( F_1^A \frac{M_+-M_--q^2}{2M_2} - F_2^A \frac{Q_+ M_-}{2M_1M_2}
+ F_3^A \frac{{|\bf p_2|^2}}{M_2} \Big)
\nn[1.1ex]
H_{\frac12 1}^A &=& \sqrt{\frac{Q_-}{q^2}} \,
\Big( F_1^A - F_2^A \frac{Q_+}{M_1M_2} \Big)
\qquad
H_{\frac32 1}^A = -  \, \sqrt{Q_-} \, F_1^A
\nonumber
\ena
We use the abbreviations
$M_\pm = M_1 \pm M_2$,
$Q_\pm = M_\pm^2 - q^2$,
${|\bf p_2|} = \lambda^{1/2}(M_1^2,M_2^2,q^2)/(2M_1)$.

For the decay width one finds

\bea
\Gamma(\Lambda_b \to \Lambda^\ast\,+\,V)
&=& \frac{G_F^2}{32 \pi} \, \frac{|{\bf p_2}|}{M_1^2} \,
|V_{cb} V^\ast_{cs}|^2 \, C_{\rm eff}^2 \, f_V^2 \, M_V^2 \,{\cal H}_N
\label{eq:LbLV_width}\\
{\cal H}_N &=& \sum_{\lambda_2,\lambda_V}|H_{\lambda_2,\lambda_V}|^2
\ena
The sum over helicities includes all helicities satisfying the angular
momentum constraint $|\lambda_2 - \lambda_v| \le 1/2$.
Compared to Eq.~(11) of~\cite{Gutsche:2013oea} we have dropped a factor
containing the lepton mass in the rate expression.

Using the helicity amplitudes one can write down angular decay distributions
in the cascade decays
$\Lambda_b\to \Lambda^{(\ast)} (\to B+M) + J/\psi(\to \ell^+\ell^-)$ where
$B$ and $M$ are the final baryon ($N$, $\Sigma$, etc.) and meson
($\pi$, $K$, etc.) states. Note that the decays $\Lambda^\ast \to p K^-$ are
strong and therefore parity conserving while the decay $\Lambda \to p \pi^-$
is a weak decay and therefore parity violating. The angular decay distribution
involving the strong decays $\Lambda^\ast \to p K^-$ can be obtained from
that involving the weak decay $\Lambda \to p \pi^-$ by setting the
relevant asymmetry parameter to zero. When the $\Lambda_b$ is polarized the angular
decay distributions
are characterized by three polar angles and two azimuthal angles.
The full five-fold angular decay distribution can be found
in~\cite{Lednicky:1985zx,Bialas:1992ny,Kadeer:2005aq}. Corresponding three-fold
polar angle distributions for
polarized $\Lambda_{b}$ decay and a threefold joint decay distribution
for unpolarized $\Lambda_{b}$ decay can be obtained from the full five-fold
decay distributions written down in~\cite{Bialas:1992ny,Kadeer:2005aq} by
appropriate angular integrations or by setting the polarization of the
$\Lambda_{b}$ to zero.

As mentioned in the introduction there are two variants of how the
angular decay distributions of such cascade decay processes can be presented.
The unprocessed form the angular decay distribution
$W(\Omega_1,\,\Omega_2,\,\theta)$ is
written down directly from the traces of the production and the rotated
decay spin density matrices. In the present case $\Omega_1$ describes the
relative orientation of
the decay $\Lambda \to p \pi^- $ (or $\Lambda^\ast \to p K^- $), $\Omega_2$ the
relative orientation of the leptonic decay $J/\psi \to \ell^+ \ell^-$
and $\theta$ the polar orientation
of the polarization of the $\Lambda_b$. In the normal form one
subtracts off unity corresponding to the normalized total rate, i.e. one writes
\be
W(\Omega_1,\,\Omega_2,\,\theta)= 1 + \widetilde W(\Omega_1,\,\Omega_2,\,\theta)
\en
where $\int d\Omega_1 d\Omega_2 d\cos\theta  \,\widetilde W(\Omega_1,\,\Omega_2,\,\theta) =0$.
The normal form of the threefold angular decay distribution
 $\widetilde W(\theta_1,\,\theta_2,\,\theta)$
can be written in terms of
three linear combinations of
normalized squared helicity amplitudes
$ |\widehat H_{\lambda_2\lambda_V}|^2$ which are~\cite{Aaij:2013oxa,Gutsche:2013oea}
\bea
\alpha_b &=&
  |\widehat H_{+\tfrac12 0}|^2 - |\widehat H_{-\tfrac12 0}|^2
+ |\widehat H_{-\tfrac12 -1}|^2 - |\widehat H_{+\tfrac12 +1}|^2
- \left(|\widehat H_{-\tfrac32 -1}|^2 - |\widehat H_{+\tfrac32 +1}|^2\right) \,,
\label{3/2}
\\
r_0 &=& |\widehat H_{+\tfrac12 0}|^2 + |\widehat H_{-\tfrac12 0}|^2 \,,\qquad
r_1  = |\widehat H_{+\tfrac12 0}|^2 - |\widehat H_{-\tfrac12 0}|^2 \,,
\label{eq:asym-param}
\ena
where $ |\widehat H_{\lambda_2\lambda_V}|^2=| H_{\lambda_2\lambda_V}|^2/{\cal H}_N$.
The last bracketed contribution in~(\ref{3/2}) only comes in for the
$1/2^+ \to 3/2^\pm$ case.


\section{The \boldmath{$\Lambda_b\to\Lambda^{(\ast)}$} form factors in the
covariant quark model}
\label{sec:model_calc}


We employ generic three-quark currents to describe the $\Lambda_Q(J^P)$ states:

\bea
\Lambda_Q(J^P) &\Longrightarrow&
\epsilon_{a_1a_2a_3} \,\Gamma_1 Q_{a_1}\,\left( u_{a_2}C\Gamma_2 d_{a_3} \right).
\label{eq:3q-cur}
\ena
Here $Q=b$ or $s$, the color index is denoted by $a_i$ and
$C = \gamma^0\gamma^2$ is the charge conjugation matrix.
The Dirac matrices $\Gamma_1$ and $\Gamma_2$ are chosen in such a way
to provide the correct P-parity for the $\Lambda$-baryons. A set of
currents for the flavor-antisymmetric $[ud]$ diquark
states is shown in Table~\ref{tab:diquark} by analogy with
the classification given in Ref.~\cite{Nielsen:2009uh}.

\begin{table}[ht]
\begin{center}
\caption{Currents for the $[ud]$ diquark states.}
\label{tab:diquark}
\vspace*{.25cm}
\def\arraystretch{1.5}
\begin{tabular}{|c|c|c|c|}
\hline
 state                &  current                                  &  $J^P$
\\
\hline
scalar diquark        & $ u_{a_2}^T C\gamma_5 d_{a_3} $              & $0^+$
\\
pseudoscalar diquark  & $ u_{a_2}^T C d_{a_3}$                       & $0^-$
\\
vector diquark        & $ u_{a_2}^T C\gamma_5\gamma_\mu d_{a_3} $     & $1^-$
\\
axial-vector diquark  & $ u_{a_2}^T C\gamma_\mu d_{a_3} $             & $1^+$
\\
\hline
\end{tabular}
\end{center}
\end{table}

One can then construct local three-quark currents with the appropriate
quantum numbers of the
the $\Lambda_Q (\frac12^\pm,\frac32^\pm)$ states. They are given by

\bea
\Lambda_Q^{1/2^+} &\Longrightarrow&
\epsilon_{a_1a_2a_3} \, Q_{a_1}\,
\left( u_{a_2}C\gamma_5 d_{a_3} \right) \,,
\nn
\Lambda_Q^{1/2^-} &\Longrightarrow&
\epsilon_{a_1a_2a_3} \,\gamma_5 Q_{a_1}\,
\left( u_{a_2}C\gamma_5 d_{a_3} \right) \,,
\nn
\Lambda_Q^{3/2^+} &\Longrightarrow&
\epsilon_{a_1a_2a_3} \,\gamma_5 Q_{a_1}\,
\left( u_{a_2}C\gamma_5\gamma_\mu d_{a_3} \right) \,,
\nn
\Lambda_Q^{3/2^-} &\Longrightarrow&
\epsilon_{a_1a_2a_3} \,Q_{a_1}\,
\left( u_{a_2}C\gamma_5\gamma_\mu d_{a_3} \right) \,.
\label{eq:Lambda-3q-cur}
\ena
Note that we do not employ derivative couplings in our interpolating
currents. It would be interesting to find out whether the use of derivative
couplings would change our results.

The covariant quark model employs nonlocal renditions of the
local three-quark currents in Eq.~(\ref{eq:Lambda-3q-cur}).
The nonlocal Lagrangian describing the couplings of the baryons
$\Lambda_Q$ ($Q=b,s$)
with their constituent quarks is then given by

\bea
{\cal L}^{\Lambda_Q}_{\rm int}(x) &=&
g_{\Lambda_Q} \,\bar \Lambda_Q(x)\cdot J_{\Lambda_Q}(x) + \mathrm{H.c.}\,,
\label{eq:lag_Lambda}\\
J_{\Lambda_Q}(x)  &=& \int\!\! dx_1 \!\! \int\!\! dx_2 \!\! \int\!\! dx_3 \,
F_{\Lambda_Q}(x;x_1,x_2,x_3) \,
\epsilon_{a_1a_2a_3}\,\Gamma_1 Q_{a_1}(x_1)\,
\left(u_{a_2}(x_2) \,C\Gamma_2 \, d_{a_3}(x_3)\right)\,,
\nn
F_{\Lambda_Q}(x;x_1,x_2,x_3) &=&
\delta^{(4)}\left(x-\sum\limits_{i=1}^3 w_i x_i\right)
\Phi_{\Lambda_Q}\left(\sum\limits_{i<j}(x_i-x_j)^2\right) \,,
\nonumber
\ena
where $w_i=m_i/(\sum\limits_{j=1}^3 m_j)$ and $m_i$ is the mass of the quark
placed at the space-time point $x_i$.

First of all, one has to calculate the $\Lambda_Q$ mass functions (or
self-energy functions) arising
from the interactions of the $\Lambda^{(\ast)}$ baryons with the constituent
quarks as written down in
Eq.~(\ref{eq:lag_Lambda}).
Then one can determine the coupling constants $g_{\Lambda_Q}$ by using
the so-called compositeness condition.

The Fourier-transforms of the mass functions are given by

\bea
\bar u(p',s)\tilde\Sigma_{1/2^\pm}(p',p)u(p,s) &=&
+ig^2_{1/2^\pm}\int\!\!dx\,e^{ip'x}\int\!\!dy\,e^{-ipy}
\bar u(p',s)\la 0| T\Big\{ J_{1/2^\pm}(x) \bar J_{1/2^\pm}(y)\Big\}|0\ra u(p,s)\,,
\nn[2mm]
u_\mu(p',s)\tilde\Sigma^{\mu\nu}_{3/2^\pm}(p',p)u_\nu(p,s) &=&
-ig^2_{3/2^\pm}\int\!\!dx\,e^{ip'x}\int\!\!dy\,e^{-ipy}
\bar u_\mu(p',s)
\la 0| T\Big\{ J^\mu_{3/2^\pm}(x) \bar J^\nu_{3/2^\pm}(y)\Big\}|0\ra u_\nu(p,s)\,.
\label{eq:mass}
\ena
The calculation of the Fourier-transforms of the vertex functions $\Phi$
can be done in a straightforward way by using Jacobi coordinates.
One arrives at the following expressions
\bea
\tilde\Sigma(p',p) &=& (2\pi)^4\, \delta^{(4)}(p'-p)\, \Sigma(p)\,,
\nn[5mm]
\Sigma(p) &=& \pm\, 6\,g^2
\int\!\!\frac{d^4k_1}{(2\pi)^4i}\int\!\!\frac{d^4k_2}{(2\pi)^4i}
\widetilde\Phi^2\left[-K^2\right]
\nn[2mm]
&\times&
\Gamma_1\, S_Q(k_1+w_1 p)\, \Gamma_1\,
\Tr\left[\, \Gamma_2\,S_u(k_2-w_2p)\, \Gamma_2\, S_d(k_2-k_1+w_3 p)\, \right] \,,
\nn[5mm]
K^2 &\equiv& \frac12 (k_1-k_2)^2 + \frac16 (k_1+k_2)^2
\label{eq:mass-2}
\ena
where   the $''+''$ sign stands for the final baryon states with $J^P=\frac12^+$
and $\frac32^-$
and the $''-''$ sign stands for the final baryon  states with $J^P=\frac12^-$
and $\frac32^+$.
We have omitted some unnecessary indices and self-explanatory notation.

In the numerical calculations we choose a simple Gaussian form
for the vertex functions (for both mesons and baryons):
\be
\tilde\Phi(-P^2) = \exp(P^{\,2}/\Lambda^2) \equiv \exp(s\,P^2)\,,
\label{eq:Gauss}
\en
where $\Lambda$ is a size parameter describing the distribution
of the quarks inside a given hadron and $s\equiv 1/\Lambda^2$.
We emphasize
that the Minkowskian momentum variable $P^{\,2}$ turns into the Euclidean form
$-\,P^{\,2}_E$ needed for the appropriate falloff behavior of the
correlation function~(\ref{eq:Gauss}) in the Euclidean region.

The compositeness condition implies that the renormalization constant of
the hadron wave function is set equal to zero.
This condition has been suggested by Weinberg~\cite{Weinberg:1962hj}
and Salam~\cite{Salam:1962ap} (for a review, see~\cite{Hayashi:1967hk})
and extensively used in our approach (for details, see~\cite{Efimov:1993ei}).
In the $J=1/2$ case the compositeness condition may be written in the form
\be
Z_{1/2} = 1 - \Sigma'_{1/2}(\not\! p)=0 \,, \qquad \not\! p = M.
\label{eq:1/2-Z=0}
\en
where $\Sigma'_{1/2}(\not\! p)$ is the derivative of the mass function taken
on the mass shell $p^2=M^2$.
In the $J=3/2$ case one has to account for the Rarita-Schwinger conditions
$p^\alpha u_\alpha(p,s)=0$ and $\gamma^\alpha u_\alpha(p,s)=0$. This can be done
by splitting off a scalar function in the form
$\Sigma_{3/2}^{\mu\nu}(p)=g^{\mu\nu}\Sigma_{3/2}(\not\! p)$.
The compositeness condition for the $J=3/2$ case reads
\be
Z_{3/2} = 1 - \Sigma_{3/2}^\prime(\not\! p)   = 0 \,, \qquad \not\! p = M.
\label{eq:3/2-Z=0}
\en
In practice, it is more convenient to use a form equivalent
to Eqs.~(\ref{eq:1/2-Z=0}) and (\ref{eq:3/2-Z=0}) by writing
\be
\frac{d\Sigma(\not\! p) }{dp^\alpha}   = \gamma^\alpha \,,\qquad
p^\alpha = M\gamma^\alpha,\quad\text{and}\quad \not\! p = M.
\label{eq:Z=0}
\en

The loop integrations in Eq.~(\ref{eq:mass-2}) are performed by using
the Fock-Schwinger representations of the quark propagators.
The tensorial loop integrations and the manipulations
with Dirac matrices are performed with the help of
FORM \cite{Vermaseren:2000nd}.
The final relations needed for the determination of the coupling constants
may be symbolically cast in the form

\bea
g &=& 1/\sqrt{ G(M,s;m_q) }\,,
\label{eq:coupling}\\
G(M,s;m_q) &=& \int\limits_0^{1/\lambda^2}\!\!dt\,t^2\!
\int\!\! d^3\alpha\,\delta\left(1-\sum\limits_{i=1}^3 \alpha_i\right)
\tilde G(t\alpha_1,t\alpha_2,t\alpha_3;M,s,m_q)
\nn[2mm]
&=&  \int\limits_0^{1/\lambda^2}\!\!dt\,t^2\!
     \int\limits_0^1\!\! d^2x\, x_1
     \tilde G(t\alpha_1,t\alpha_2,t\alpha_3;M,s,m_q)\,,
\nn[2mm]
&&
\alpha_1=1-x_1,\quad \alpha_2=x_1(1-x_2),\quad \alpha_3=x_1x_2.
\nonumber
\ena
The infrared cutoff parameter $\lambda$ provides for
the absence of all constituent quark threshold singularities.
The threefold integrals are calculated by a FORTRAN code using
the NAG library.

The matrix elements of the transitions
$\la \Lambda_2 | \bar s \Gamma_\mu b | \Lambda_1 \ra $
finally read

\bea
\la \Lambda_2 |\, \bar s\, \Gamma_\mu b\, | \Lambda_1 \ra
&=& 6\,g_{\Lambda_1} g_{\Lambda_2}
\int\!\!\frac{d^4k_1}{(2\pi)^4i}\int\!\!\frac{d^4k_2}{(2\pi)^4i}
\widetilde\Phi_{\Lambda_1}\left[-\Omega_1^2\right]
\widetilde\Phi_{\Lambda_2}\left[-\Omega_2^2\right]
\nn[2mm]
&\times&
\bar u_2(p_2,s_2)\,\Gamma_1\,S_s(k_1+p_2)\,\Gamma_\mu \,S_b(k_1+p_1)\,
\Tr\left[\, S_u(k_2)\,\Gamma_2\,S_d(k_2-k_1)\,\gamma_5 \right]\,u_1(p_1,s_1)\,,
\nn[5mm]
\Omega_1^2 &\equiv&  \tfrac12 (k_1-k_2 + v_3\, p_1)^2
                   + \tfrac16 (k_1+k_2+(2\,v_2+v_3)\, p_1 )^2\,,
\nn[2mm]
\Omega_2^2 &\equiv&  \tfrac12 (k_1-k_2 + w_3\, p_2)^2
                   + \tfrac16 (k_1+k_2+(2\,w_2+w_3) p_2 )^2\,.
\label{eq:L1L2}
\ena
where $\Gamma_\mu=\gamma_\mu$ or  $\Gamma_\mu=\gamma_\mu\gamma_5$,
$\Lambda_1 = \Lambda_b(p_1,s_1)$ and
$\Lambda_2 = \Lambda^{(\ast)}(p_2,s_2)$.
The reduced quark masses are defined by
\bea
v_1 &=& \frac{m_b}{m_{bud}} \,, \qquad
v_2  =  \frac{m_u}{m_{bud}}\,, \qquad
v_3  =  \frac{m_d}{m_{bud}}\,,  \qquad m_{bud}=m_b+m_u+m_d\,,
\nn[2mm]
w_1 &=& \frac{m_s}{m_{sdu}}\,, \qquad
w_2  =  \frac{m_u}{m_{sdu}}\,, \qquad
w_3  =  \frac{m_d}{m_{sdu}}\,,\qquad m_{sdu} = m_s+m_d+m_u \,.
\nonumber
\ena

Below we show the different Dirac structures $\Gamma_1$ and $\Gamma_2$
in~(\ref{eq:L1L2}) 
for the different final state $\Lambda^{(\ast)}$ baryons:

\[
\begin{array}{lr||lr}
\hline
J^P \qquad & \qquad \Gamma_1\otimes\Gamma_2       \qquad & \qquad
J^P \qquad & \qquad \Gamma_1\otimes\Gamma_2
\\
\hline
&&&\\[-3mm]
\frac12^+ \qquad & \qquad I\otimes\gamma_5        \qquad & \qquad
\frac32^+ \qquad & \qquad \gamma_5\otimes\gamma_5\gamma_\alpha
\\[2mm]
\frac12^- \qquad & \qquad \gamma_5\otimes\gamma_5   \qquad & \qquad
\frac32^- \qquad & \qquad I\otimes\gamma_5\gamma_\alpha
\\[2mm]
\hline
\end{array}
\]

 The expressions for the scalar form factors are represented
by the fourfold integrals

\bea
&&
F(M_1,s_1,M_2,s_2,m_q,q^2) =
6\,g_{\Lambda_1}\,g_{\Lambda_2}\! \int\limits_0^{1/\lambda^2}\!\!dt\,t^3\!
     \int\limits_0^1\!\! d^3x\, x_1^2\,x_2
     \tilde F(t\alpha_1,t\alpha_2,t\alpha_3,t\alpha_4;M_1,s_1,M_2,s_2,m_q,q^2)\,,
\nn[2mm]
&&
\alpha_1=1-x_1,\quad \alpha_2=x_1(1-x_2),\quad \alpha_3=x_1x_2(1-x_3), \quad
\alpha_4=x_1x_2x_3\,.
\nonumber
\ena

The model parameters are the constituent quark masses $m_q$ and
the infrared cutoff parameter $\lambda$ responsible for quark confinement.
They are taken from a new fit done and used in
our  papers on the semileptonic $B\to D^{(\ast)}\ell\bar\nu_\ell$
decays~\cite{Ivanov:2017mrj,Ivanov:2016qtw,Ivanov:2015woa,Ivanov:2015tru},
rare $B\to M\bar\ell\ell$
decays~\cite{Issadykov:2015iba,Dubnicka:2015iwg,Dubnicka:2016nyy},
the semileptonic decays
$\Lambda_b \to \Lambda_c + \tau^- + \bar{\nu_\tau}$,
$\Lambda_c^+ \to \Lambda \ell^+ \nu_\ell$~\cite{Gutsche:2015mxa,Gutsche:2015rrt}
and for the calculation of nucleon tensor form factors~\cite{Gutsche:2016xff}.
The best fit values for the constituent quark masses and the infrared cutoff
parameter $\lambda$ are
\be
\def\arraystretch{2}
\begin{array}{ccccccc}
     m_u        &      m_s        &      m_c       &     m_b & \lambda  &
\\\hline
 \ \ 0.241\ \   &  \ \ 0.428\ \   &  \ \ 1.67\ \   &  \ \ 5.05\ \   &
\ \ 0.181\ \   & \ {\rm GeV}
\end{array}
\label{eq: fitmas}
\en

The dimensional-size parameters of the ground-state $\Lambda_b$ and
$\Lambda_s$ baryons
have been determined  by a fit to the semileptonic
decays $\Lambda_{b} \to \Lambda_{c}+ \ell^{-}\bar \nu_{\ell}$ and
$\Lambda_{c} \to \Lambda+ \ell^{+} \nu_{\ell}$. The resulting values are
$\Lambda_{\Lambda_b} = 0.571$,
$\Lambda_{\Lambda_s} = 0.492$~GeV. The values
of the size parameters of the final states $\Lambda^\ast(\frac12^-,\frac32^\pm)$
are set equal to the size parameter of the ground state $\Lambda_{\Lambda_s}$.
For the size parameter of the $J/\psi$ we take $\Lambda_{J/\psi} = 1.74$~GeV
as determined from our most recent fit described above.


\section{Numerical results}
\label{sec:numerics}

In Tables~\ref{tab:hel} and \ref{tab:branching} we list our predictions for
the normalized helicity amplitudes and branching fractions.

\begin{table}[htb]
\begin{center}
\caption{Moduli squared of normalized helicity amplitudes.}
\label{tab:hel}
\vspace*{.1cm}
\def\arraystretch{1.7}
\begin{tabular}{|c|cccc|}
\hline
$\Lambda^\ast$ & 1116 & 1405 & 1890 & 1520 \\
\hline
$J^P$         & $\frac12^+$ & $\frac12^-$ &   $\frac32^+$ &  $\frac32^-$
\\
\hline
\qquad $|\hat H_{+\tfrac32 +1}|^2$ \qquad  &\qquad 0  \qquad &\qquad 0 \qquad
&\qquad $3.50\times 10^{-4}$\qquad &\qquad $0.84\times 10^{-4}$ \qquad
\\
\qquad $|\hat H_{+\tfrac12 +1}|^2$\qquad  &\qquad $2.34\times 10^{-3}$\qquad
&\qquad $1.27\times 10^{-2}$\qquad  &\qquad $3.19\times 10^{-2}$\qquad
&\qquad $2.26\times 10^{-2}$\qquad
\\
\qquad $|\hat H_{+\tfrac12 0}|^2$\qquad   &\qquad $3.24\times 10^{-4}$\qquad
&\qquad $5.19\times 10^{-3}$\qquad  &\qquad $1.61\times 10^{-3}$\qquad
&\qquad $1.82\times 10^{-3}$\qquad
\\
\qquad $|\hat H_{-\tfrac12 0}|^2$\qquad   &\qquad 0.53 \qquad &\qquad 0.51\qquad
&\qquad $0.51$\qquad        &\qquad 0.54\qquad
\\
\qquad $|\hat H_{-\tfrac12 -1}|^2$\qquad  &\qquad 0.47 \qquad &\qquad 0.47 \qquad
&\qquad $ 0.45$ \qquad       & \qquad 0.44 \qquad
\\
\qquad $|\hat H_{-\tfrac32 -1}|^2$\qquad  &\qquad 0 \qquad &\qquad 0 \qquad
&\qquad $3.34\times 10^{-3}$\qquad  &\qquad $1.06\times 10^{-3}$ \qquad 
\\[2mm]
\hline
\end{tabular}
\end{center}
\end{table}
The helicity amplitudes $H_{\lambda_2,\lambda_V}$ of the produced $\Lambda^{(*)}$
states are clearly dominated by
the helicity configuration $\lambda_2=-1/2$ as in the quark level transition
$b \to s$. For the spin $1/2$ states in the
transition $1/2^+ \to 1/2^\pm$ this implies that the two $\Lambda^{(*)}(1/2)$
states are almost purely left-handed.

\begin{table}[htb]
\begin{center}
\caption{Branching ratio ${\cal B}(\Lambda_b\to \Lambda^\ast + J/\psi$) (in units of $10^{-4}$).}
\label{tab:branching}
\vspace*{.1cm}
\def\arraystretch{1.7}
\begin{tabular}{|c|cccc|}
\hline
$\Lambda^\ast$ & 1116 & 1405 & 1890 & 1520 \\
\hline
$J^P$         & $\frac12^+$ & $\frac12^-$ &   $\frac32^+$ &  $\frac32^-$ \\
\hline
${\cal B}\times 10^4$    &  8.00      & 7.07        & 0.45          & 0.19   \\[1.5mm]

\hline
\end{tabular}
\end{center}
\end{table}
It is also apparent that the branching ratios involving the excited
$J^P=3/2^\pm$ states
are suppressed relative to those of the ground state $\Lambda(1116)$ and the
excited state with $J^P=1/2^-$.

\begin{table}[!htbp]
\caption{Asymmetry parameters and moduli squared of normalized helicity amplitudes $\tH{\lambda_\varLambda\lambda_\psi}$
  for the $\varLambda_b^0 \to \varLambda^0$ transition.}
\begin{tabularx}{0.9\textwidth}{|r|R|R|R|R|}
\hline
$\varLambda^{(\star)}$, $J^P$ & \multicolumn{4}{c|}{$\varLambda$, $\tfrac12^+$} \\
\hline
Quantity & \multicolumn{1}{c|}{Our results} & \multicolumn{1}{c|}{LHCb~\cite{Aaij:2013oxa}}
                                            & \multicolumn{1}{c|}{ATLAS~\cite{Aad:2014iba}}
                                            & \multicolumn{1}{c|}{CMS~\cite{CMS:2016iaf}} \\
\hline
$\tH{+\tfrac12 +1}$ & 2.34\times 10^{-3} & -0.10 \pm 0.04 \pm 0.03 & (0.08^{+0.13}_{-0.08} \pm 0.06)^2 &  0.05 \pm 0.04 \pm 0.02 \\
$\tH{+\tfrac12  0}$ & 3.24\times 10^{-4} &  0.01 \pm 0.04 \pm 0.03 & (0.17^{+0.12}_{-0.17} \pm 0.09)^2 & -0.02 \pm 0.03 \pm 0.02 \\
$\tH{-\tfrac12  0}$ & 0.532             &  0.57 \pm 0.06 \pm 0.03 & (0.59^{+0.06}_{-0.07} \pm 0.03)^2 &  0.51 \pm 0.03 \pm 0.02 \\
$\tH{-\tfrac12 -1}$ & 0.465             &  0.51 \pm 0.05 \pm 0.02 & (0.79^{+0.04}_{-0.05} \pm 0.02)^2 &  0.46 \pm 0.02 \pm 0.02 \\[1ex]
\hline
$\alpha_b$ & -0.069 &  0.05 \pm 0.17 \pm 0.07 & 0.30 \pm 0.16 \pm 0.06 & -0.12 \pm 0.13 \pm 0.06 \\
$r_0$      &  0.533 &  0.58 \pm 0.02 \pm 0.01 &                        & \\
$r_1$      & -0.532 & -0.56 \pm 0.10 \pm 0.05 &                        & \\
\hline
\end{tabularx}
\label{tab:tH12_Lambda0}
\end{table}

\begin{table}[!htbp]
\caption{Asymmetry parameters and moduli squared of normalized helicity amplitudes $\tH{\lambda_{\varLambda^\star}\lambda_\psi}$
for the $\varLambda_b^0 \to \varLambda^\star(\tfrac12^\pm)$ transition.}
\begin{tabularx}{0.9\textwidth}{|r|R|R|R|R|R|R|R|R|}
\hline
$\varLambda^{(\star)}$, $J^P$
                          & \multicolumn{2}{c|}{$\varLambda(1405)$, $\tfrac12^-$}
                          & \multicolumn{2}{c|}{$\varLambda(1600)$, $\tfrac12^+$}
                          & \multicolumn{2}{c|}{$\varLambda(1800)$, $\tfrac12^-$}
                          & \multicolumn{2}{c|}{$\varLambda(1810)$, $\tfrac12^+$} \\
\hline
Quantity & \multicolumn{1}{c|}{Our results} & \multicolumn{1}{c|}{LHCb~\cite{Jurik:2016bdm}}
         & \multicolumn{1}{c|}{Our results} & \multicolumn{1}{c|}{LHCb~\cite{Jurik:2016bdm}}
         & \multicolumn{1}{c|}{Our results} & \multicolumn{1}{c|}{LHCb~\cite{Jurik:2016bdm}}
         & \multicolumn{1}{c|}{Our results} & \multicolumn{1}{c|}{LHCb~\cite{Jurik:2016bdm}} \\
\hline
$\tH{+\tfrac12 +1}$ & 1.27\times 10^{-2} &  0.025 & 4.08\times 10^{-2} &  0.105 & 2.33\times 10^{-2} &  0.137 & 6.90\times 10^{-2} &  0.059 \\
$\tH{+\tfrac12  0}$ & 5.19\times 10^{-3} &  0.241 & 1.05\times 10^{-2} &  0.085 & 5.23\times 10^{-3} &  0.176 & 1.99\times 10^{-2} &  0.243 \\
$\tH{-\tfrac12  0}$ & 0.514             &  0.143 & 0.458             &  0.455 & 0.457             &  0.422 & 0.418             &  0.478 \\
$\tH{-\tfrac12 -1}$ & 0.468             &  0.592 & 0.491             &  0.345 & 0.514             &  0.266 & 0.493             &  0.221 \\[1ex]
\hline
$\alpha_b$          & -0.054            &  0.665 &  0.003            & -0.130 &  0.039            & -0.117 &  0.026            & -0.073 \\
$r_0$               &  0.519            &  0.384 &  0.468            &  0.540 &  0.462            &  0.598 &  0.438            &  0.721 \\
$r_1$               & -0.509            &  0.098 & -0.447            & -0.370 & -0.452            & -0.246 & -0.398            & -0.235 \\
\hline
\end{tabularx}
\label{tab:tH12}
\end{table}

\begin{table}[!htbp]
\caption{Asymmetry parameters and moduli squared of normalized helicity amplitudes $\tH{\lambda_{\varLambda^\star}\lambda_\psi}$
for the $\varLambda_b^0 \to \varLambda^\star(\tfrac32^\pm)$ transition.}
\begin{tabularx}{0.9\textwidth}{|r|R|R|R|R|R|R|}
\hline
$\varLambda^\star$, $J^P$ & \multicolumn{2}{c|}{$\varLambda(1520)$, $\tfrac32^-$}
                          & \multicolumn{2}{c|}{$\varLambda(1690)$, $\tfrac32^-$}
                          & \multicolumn{2}{c|}{$\varLambda(1890)$, $\tfrac32^+$} \\
\hline
Quantity & \multicolumn{1}{c|}{Our results} & \multicolumn{1}{c|}{LHCb~\cite{Jurik:2016bdm}}
         & \multicolumn{1}{c|}{Our results} & \multicolumn{1}{c|}{LHCb~\cite{Jurik:2016bdm}}
         & \multicolumn{1}{c|}{Our results} & \multicolumn{1}{c|}{LHCb~\cite{Jurik:2016bdm}} \\
\hline
$\tH{+\tfrac32 +1}$ & 8.37\times 10^{-5} &  0.067 & 2.44\times 10^{-4} &  0.054 & 3.50\times 10^{-4} &  0.297 \\
$\tH{+\tfrac12 +1}$ & 2.26\times 10^{-2} &  0.107 & 4.67\times 10^{-2} &  0.031 & 3.19\times 10^{-2} &  0.130 \\
$\tH{+\tfrac12  0}$ & 1.82\times 10^{-3} &  0.047 & 4.98\times 10^{-3} &  0.492 & 1.61\times 10^{-3} &  0.236 \\
$\tH{-\tfrac12  0}$ & 0.536             &  0.552 & 0.509             &  0.257 & 0.512             &  0.078 \\
$\tH{-\tfrac12 -1}$ & 0.439             &  0.109 & 0.437             &  0.040 & 0.451             &  0.207 \\
$\tH{-\tfrac32 -1}$ & 1.06\times 10^{-3} &  0.119 & 1.78\times 10^{-3} &  0.126 & 3.34\times 10^{-3} &  0.053 \\[1ex]
\hline
$\alpha_b$          & -0.118            & -0.555 & -0.115            &  0.172 & -0.094            &  0.479 \\
$r_0$               &  0.537            &  0.599 &  0.514            &  0.749 &  0.514            &  0.314 \\
$r_1$               & -0.534            & -0.505 & -0.504            &  0.235 & -0.510            &  0.158 \\
\hline
\end{tabularx}
\label{tab:tH32}
\end{table}

\begin{figure}[H]
\begin{tabular}{cc}
  \includegraphics[width=0.50\linewidth]{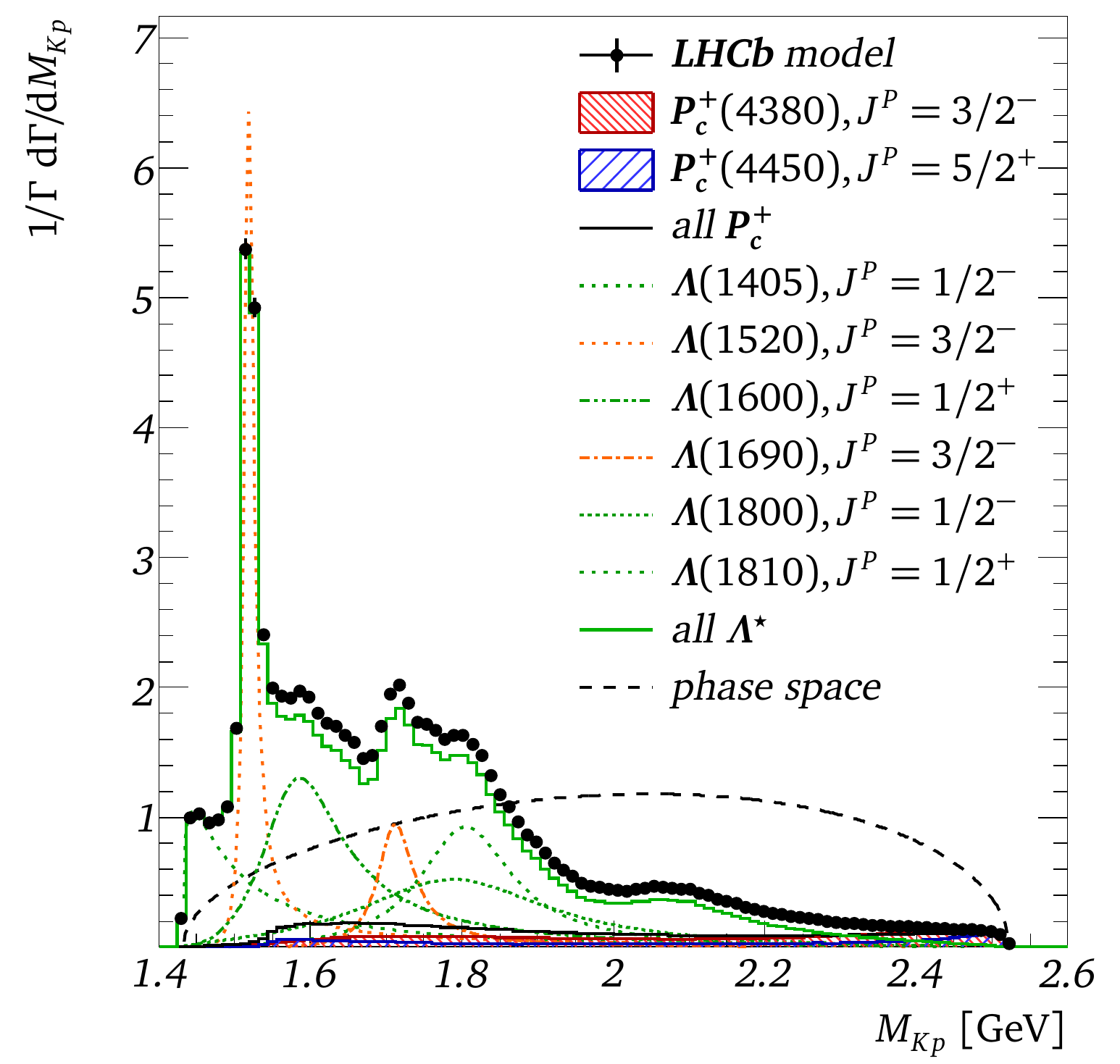} &
  \includegraphics[width=0.50\linewidth]{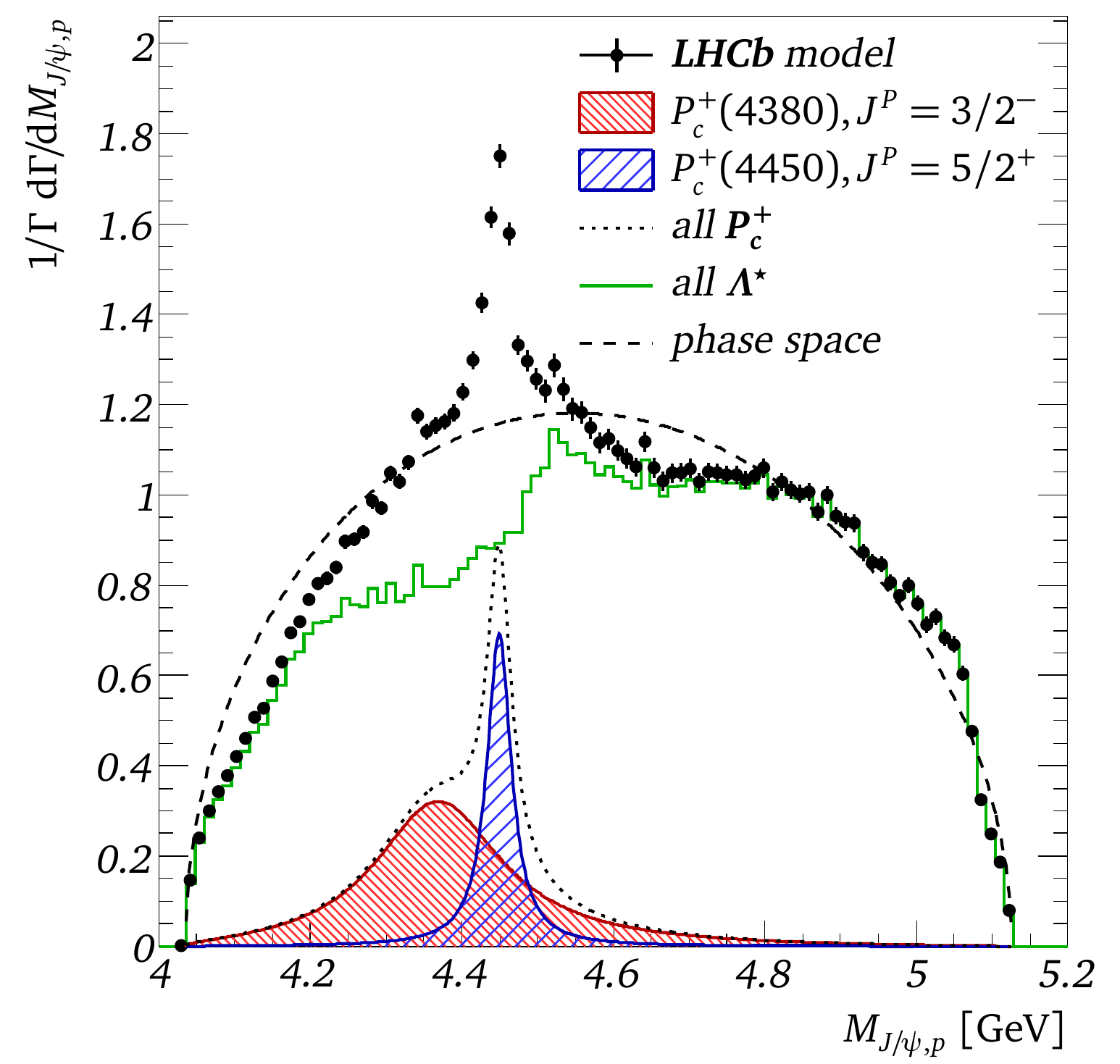}
\end{tabular}
\caption{The invariant mass $M(Kp)$ (left) and $M(\Jpsi p)$ (right)
  distributions. Full LHCb model pseudodata are shown as black dots with
  error bars, while the hatched area corresponds to the $P_c^+$ exotic states.
  The main contributions from the $\varLambda^{(*)}$ resonances are also shown. }
\label{fig:mLambdaStar}
\end{figure}

\begin{figure}[H]
\begin{tabular}{cc}
  \includegraphics[width=0.50\linewidth]{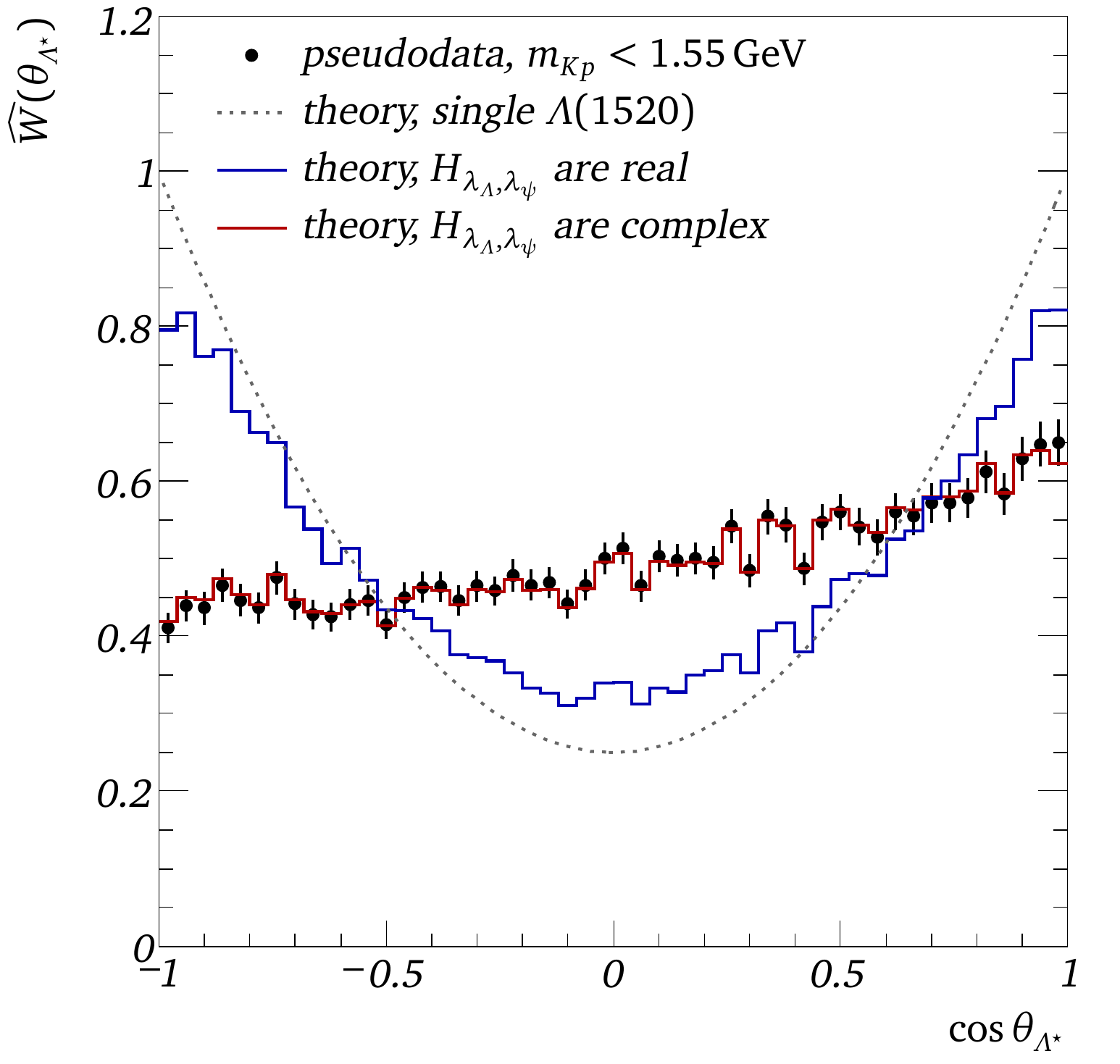} &
  \includegraphics[width=0.50\linewidth]{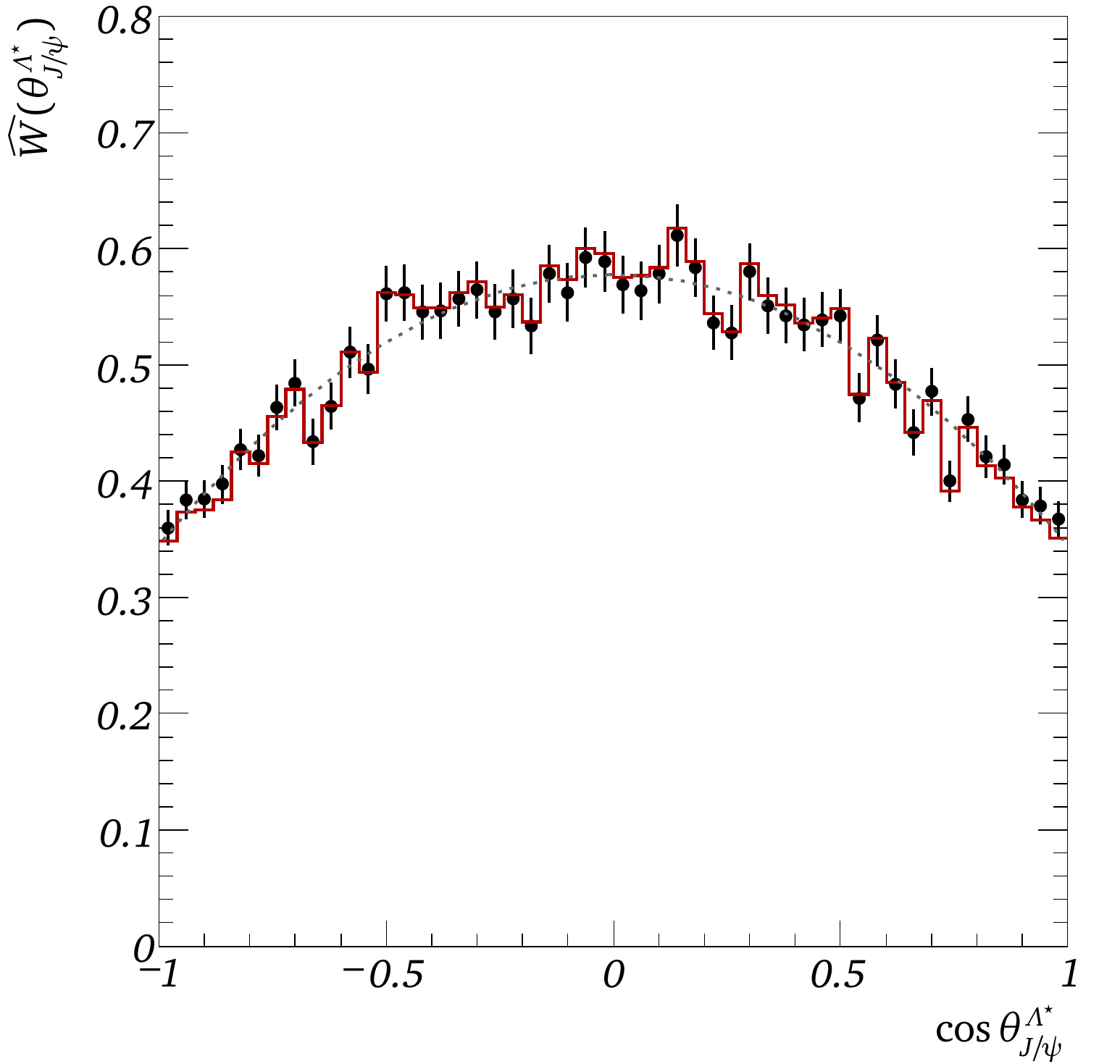}
\end{tabular}
\caption{Helicity angles $\theta_{\varLambda^\star}$ (left) and $\theta_{\Jpsi}$ (right) distributions for the low mass $M_{Kp}$ region
  ($M_{Kp}<1.55~\text{GeV}$). Full LHCb model (pseudodata) is shown as black dots with error bars, while the dotted line represent our
  calculation for the single $\varLambda(1520)$ state. Solid red and blue lines correspond to our attempt to describe the pseudodata
  by taking into account only the 3 lowest $\varLambda^\star$ states, see details in text.}
\label{fig:theta}
\end{figure}

Next, we discuss the experimental results on the pentaquark evidence in 
light of our theoretical findings.
Large data samples of the bottom baryon state $\varLambda_b^0$ were
collected by
the LHC experiments
from $pp$ collisions during Run I. An angular analysis of the decay
$\varLambda_b^0 \to \Jpsi(\mu^+\mu^-)\varLambda^0(p\pi^-)$
was first done by the LHCb Collaboration.~\cite{Aaij:2013oxa} and was 
then repeated by the ATLAS~\cite{Aad:2014iba} and CMS~\cite{CMS:2016iaf} 
Collaborations.
From~Table~\ref{tab:tH12_Lambda0}, one can assess the present-day accuracy of
the measured helicity amplitudes
for the transition $\varLambda_b^0 \to \varLambda^0(J^P=1/2^+)$.

The dominant production mechanism of the
$\varLambda_b^0$'s at the LHC proceeds via the strong interactions. Therefore
the longitudinal polarization $P_L$ of the produced $\varLambda_b^0$
vanishes because of parity conservation.
Contrary to this a transverse polarization component $P_T$ is not
forbidden by parity. The $P_T$ component depends strongly on the Feynman
variable $x_F = 2p_\parallel/\sqrt{s}$, where
$p_\parallel$ is the longitudinal momentum of the $\varLambda_b^0$
and $\sqrt{s}$ is the collision center-of-mass energy.
For the collisions of identical unpolarized initial state particles one has
$P_T(-x_F) = - P_T(x_F)$
by virtue of the invariance under the rotation of the coordinate system through an angle of $180^\circ$ about
the normal $\vec{n}$ to the reaction plane~\cite{Abramov:2005sb}.
This implies that $P_T(x_F = 0) = 0$.
Taking into account the very small value $x_F \approx 0.02$ for
$\varLambda_b^0$'s produced at the LHC
at $\sqrt{s}=7$~TeV, $P_T$ is estimated to be $\mathcal{O}(10\%)$
in~\cite{Hiller:2007ur}.
The $P_T$ value measured in~\cite{Aaij:2013oxa,Aad:2014iba,CMS:2016iaf} is
consistent with the expected value of zero.
We therefore treat $\varLambda_b^0$ to be unpolarized in our further
analysis.

Large samples of the decay $\varLambda_b^0 \to \Jpsi K^- p$ decay have been
collected by the LHCb experiment~\cite{Aaij:2015tga}.
This decay was expected to be dominated by $\varLambda^\star$ resonances
decaying into $K^- p$ final states.
The measured fit fractions of the $\varLambda(1405)$ and $\varLambda(1520)$
states are $(15\pm 1\pm 6)\%$ and $(19\pm 1\pm 4)\%$,
respectively.
It was also found that the data cannot be satisfactorily described without
the inclusion of two Breit-Wigner resonances
decaying strongly to $\Jpsi\, p$. These new pentaquark states,
called $P_c(4380)^+$ and
$P_c(4450)^+$, have large fit fractions
of $(4.1\pm 0.5 \pm 1.1)\%$ and $(8.4\pm 0.7 \pm 4.2)\%$ of the total
$\varLambda_b^0 \to \Jpsi K^- p$ sample, respectively.

The presence of various conventional $\varLambda^\star \to K^- p$ resonances
and exotic pentaquark states $P_c^+\to \Jpsi p$,
which can interfere with each other, makes the analysis of experimental
data particularly difficult and challenging.
There are 32 complex parameters ($B_{L,S}$ amplitudes) which describe the
transition $\varLambda_b^0 \to \varLambda^\star$. The 32 parameters were
determined in a six-dimensional fit~\cite{Aaij:2015tga} to the angular
decay distribution of the cascade decay process.
Their numerical values for the default fit variant can be found in the
Appendix G of~\cite{Jurik:2016bdm}.

The helicity amplitudes
$H_{\lambda_2,\lambda_\Jpsi}(\Lambda_b \to \Lambda^{(*)} \Jpsi)$
used in the present approach
are linearly related to the $LS$-amplitudes $B_{LS}$  used
in~\cite{Aaij:2015tga}(see e.g.~\cite{ms1970,chung2014}). The coefficients 
of this linear relation can be obtained with the help
of angular momentum Clebsch-Gordan coefficients
(see Eq.~(2) in~\cite{Aaij:2015tga}).
There are two important remarks that one has to make here.
First, the $B_{LS}$ amplitudes from the fit have already been redefined in order
to account for the helicity couplings from the strong decays
$\Lambda^{*} \to K^- p$, i.e. the experiment reports values for the
products
$H_{\lambda_2,\lambda_{J/\psi}}(\Lambda_b^0 \to \Lambda^{*} J/\psi)
H_{+1/2}(\Lambda^{*} \to K^- p)$. It is obvious that the additional factor
$ H_{+1/2}(\varLambda_n^\star \to K^- p)$ still allows one to compare the moduli 
of normalized helicity amplitudes with theoretical expectations.
The second remark concerns the unusual sign of the helicity $\lambda_p$ in
the $\varLambda^\star$ decay chain in~\cite{Jurik:2016bdm}
leading to a redefinition
$H_{\lambda_\varLambda^\star,\lambda_{\Jpsi}}(\varLambda_b^0 \to \varLambda_n^\star \Jpsi) \to
(\pm) (H_{-\lambda_\varLambda^\star,-\lambda_{\Jpsi}}(\varLambda_b^0 \to \varLambda_n^\star \Jpsi)^*$.

In Tables~\ref{tab:tH12} and~\ref{tab:tH32}, we compare our predictions for 
the moduli squared of normalized helicity amplitudes
with values recalculated from a fit to experimental data~\cite{Jurik:2016bdm}.
Unfortunately, the absence of the correlation matrix does not allow us to estimate the error bands for those values;
nevertheless, they are expected to be rather large.
We found that the predicted values of $H_{\lambda_\varLambda^\star,\lambda_\Jpsi}$ do not vary significantly with the invariant mass $M(Kp)$ and
the leading contribution should come from the $H_{-1/2, 0}$ and $H_{-1/2, -1}$
helicity amplitudes
for both the $J^P=\tfrac12^\pm$ and $\tfrac32^\pm$ cases.
No such pattern could be found in the experimental data (may be except for $r_0$).

At this stage we cannot decide definitely whether our theoretical approach
contradicts the experimental analysis or not.
A refit of the LHCb data using our theoretical restrictions for the helicity
amplitudes is needed to obtain an unambiguous conclusion.

Despite this fact, we propose a simple check.
For this purpose we have generated a sample of 750k events containing the decay
$\varLambda_b^0 \to \Jpsi K^- p$
using the PYTHIA~8.1~\cite{Sjostrand:2007gs} Monte Carlo generator.
The decay products are distributed isotropically in the
$\varLambda_b^0$ rest frame.
We have used an event-by-event reweighting to take into account the presence
of resonances as well as possible interference effects between them.
The weight is given by the matrix element squared and is calculated for each
simulated $\varLambda_b^0$ decay
using the 4-momenta of the outgoing particles.
We have tried to reproduce the default LHCb fit which is known to describe
the experimental data well
by using the corresponding matrix elements (Eq.~(8) from~\cite{Aaij:2015tga})
with the appropriate constants given in~\cite{Jurik:2016bdm}.

The invariant mass distributions for the $M(Kp)$ and $M(\Jpsi p)$ invariant
masses are shown in Fig.~\ref{fig:mLambdaStar}.
We have concentrated on events with $M(Kp)<1.55~\text{GeV}$ where this
subsample contains about 39k of the  $\varLambda_b^0$ decays.
The contribution from the $\varLambda(1520)$ state should be dominant here
while the influence of $P_c^+$ pentaquark states can be safely neglected.

Next we have tried to consider the helicity angle distributions for $\varLambda_b^0$, $\varLambda^\star$ and $\Jpsi$
(our definitions for helicity angles are identical with those of the LHCb
analysis).
$\widehat{W}(\theta_{\varLambda_b^0})$ should be trivial for the decay of a
unpolarized $\varLambda_b^0$ particle.
The shape of the helicity angle $\theta_{\Jpsi}$ distribution is the same as for the single resonance:
$\widehat{W}(\theta_{\Jpsi}) =\tfrac12 (1+\tfrac{A}{2}(3\cos^2\theta_{\Jpsi}-1))$,
with a coefficient $A \sim (1-3r_0)/2$.

The most interesting distribution $\widehat{W}(\theta_{\varLambda^\star})$ is
shown on the left plot in Fig.~\ref{fig:theta}.
The black dotted curve corresponds to the full LHCb model. The curve does not
look
like the expected even function of the cosine of the $\varLambda^\star$
helicity angle. The main reason is the strong interference between different
intermediate $\varLambda^\star$ resonances.
We have tried to describe this distribution by taking into account only the
three lowest $\varLambda^\star$ states
(i.e. $\varLambda(1405)$, $\varLambda(1520)$ and $\varLambda(1600)$). We have
further
neglected all helicity amplitudes except for $H_{-1/2, 0}$ and $H_{-1/2, -1}$;
the overall fraction of each $\varLambda^\star$ resonance is also fixed to
its LHCb values.

In our approach all helicity amplitudes $H_{\lambda_\varLambda^\star,\lambda_\Jpsi}$ are
real.
Complex phases can
result from the decay $\varLambda^\star \to K^- p$ through final state
interactions.
To start with we assign an identical complex phase to each helicity
amplitude.
The result is shown on Fig.~\ref{fig:theta} by a solid blue line.
Obviously, it does not agree with the reference plot.
One can achieve a reasonable agreement by varying the  5 complex phases of
the different helicity amplitudes $H_{\lambda_2\,\lambda_V}$ and keeping
the moduli of the helicity amplitudes $H_{\lambda_2\,\lambda_V}$ unchanged
(see the solid red line in Fig.~\ref{fig:theta}).
We take this as evidence that experimental data can in fact be described
within our theoretical approach which includes quite strong constraints for the
moduli of the helicity amplitudes.

\section{Summary}
\label{sec:summary}

We have calculated the invariant and helicity amplitudes in
the transitions $\Lambda_b~\to~\Lambda^{(\ast)}(J^P)~+~J/\psi$ where the
$\Lambda^{(\ast)}(J^P)$ states are  $(sud)$-resonances with $J^P$ quantum numbers
$J^P=\frac12^{\pm},\frac32^{\pm}$.
The calculations were performed in the framework of a covariant
confined quark model previously developed by us.
We have found that the values of the helicity amplitudes for the transitions
into the $\Lambda^\ast(1520,\,\frac32^-)$ and $\Lambda^\ast(1890,\,\frac32^+)$
states are suppressed compared with those for the transitions into the
ground state $\Lambda(1116,\,\frac12^+)$ and the excited state
$\Lambda^\ast(1405,\,\frac12^-)$.

We have compared our numerical results for the helicity amplitudes and decay
assymmetry parameters for the set of $\Lambda^\ast$ resonances with those
recalculated from the LHCb fit.
We have shown that the helicity angle distributions for the low-mass
$\Lambda^\ast$ resonances
($M(Kp)<1.55~\text{GeV}$) can be reproduced using our predicted values
for the normalized helicity amplitudes $\widehat{H}_{-1/2, 0}$ and
$\widehat{H}_{-1/2, -1}$.

This analysis is important  for the identification of
the hidden charm pentaquark states $P_c^+(4450)$ and $P_c^+(4380)$ since the
cascade decay
$\Lambda_b~\to~\Lambda^\ast(\frac12^-,\frac32^\pm)(~\to~pK^-)~+~J/\psi$
involves the same final  states as the decay
$\Lambda_b^0~\to~P_c^+(~\to~p~K^-)~+~J/\psi $.

\begin{acknowledgments}

This work was supported
by the German Bundesministerium f\"ur Bildung und Forschung (BMBF)
under Project 05P2015 - ALICE at High Rate (BMBF-FSP 202):
``Jet- and fragmentation processes at ALICE and the parton structure
of nuclei and structure of heavy hadrons'',
by CONICYT  (Chile) PIA/Basal FB0821,
by the Tomsk State University Competitiveness
Improvement Program and the Russian Federation program ``Nauka''
(Contract No. 0.1764.GZB.2017).
M.A.I.\ acknowledges the support from  the PRISMA cluster of excellence
(Mainz Uni.). M.A.I. and J.G.K. thank the Heisenberg-Landau Grant for
the partial support. P.S. acknowledges support by the INFN (QFT-HEP project). 

\end{acknowledgments}

\end{document}